\documentclass[aps, pre, twocolumn]{revtex4-2}
\usepackage{graphicx}
\usepackage[x11names]{xcolor}

\usepackage[
  colorlinks,
  citecolor=DarkOrchid3,
  linkcolor=Cyan4,
  urlcolor=DarkOrchid3
]{hyperref}

\usepackage{amsmath, amssymb}   % for math
\usepackage{amsfonts, dsfont}
\usepackage{microtype, booktabs}

\DeclareMathOperator{\diag}{diag}

\begin{document}

\title{Measuring dynamical systems on directed hypergraphs.}
\author{Mauro Faccin}
\affiliation{Institut de Recherche pour le D\'eveloppement (IRD) and Centre Population et D\'eveloppement (CEPED), University of Paris, Boulevard Saint-Germain 75006 Paris, France.}

\begin{abstract}
  Networks and graphs provide a simple but effective model to a vast set of systems in which building blocks interact throughout pairwise interactions.
  Unfortunately, such models fail to describe all those systems in which building blocks interact at a higher order.
  Higher-order graphs provide us the right tools for the task, but introduce a higher computing complexity due to the interaction order.

  In this paper we analyze the interplay between the structure of a directed hypergraph and a linear dynamical system, a random walk, defined on it.
  How can one extend network measures, such as centrality or modularity, to this framework?
  Instead of redefining network measures through the hypergraph framework, with the consequent complexity boost, we will measure the dynamical system associated to it.
  This approach let us apply known measures to pairwise structures, such as the transition matrix, and determine a family of measures that are amenable to such a procedure.
\end{abstract}

\maketitle

\section{Introduction}%
\label{sec:introduction}

Studying complex systems often requires disaggregating those systems into simpler parts that interact with each other.

Network science often names those base constituents \emph{nodes} and the pairwise interaction between them \emph{edges}.
Unfortunately, in some cases, pairwise interactions fall short in describing the initial system and one needs to follow a different route.
Within the different methods introduced to tackle such a problem, simplicial complexes introduce the notion of nodes that interact through simplices~\cite{bianconi2016simplicial,petri2018simplicial} such as a segment, a triangle, a tetrahedron etc.
Hypergraphs, on the other hand, define interactions beyond pairwise through the definition of hyperedges, edges that affect multiple nodes at once~\cite{zhou2007learning_hypergraphs, Schaub_Signal_proc_2021,Bellaachia_2021}.
We will focus on the latter which provides a natural framework to describe a directed dynamical process.

When dynamical processes are coupled to the linking pattern of the interacting nodes, the double choice of the topology and the dynamical model will greatly influence the analysis of the original system.
In hypergraphs, coupling with dynamics~\cite{Battiston_2020_hypergraphs_dynamics} comes in a wide variety of flavors with different purposes.
Synchronization of oscillators may display chimera states only in higher order topologies~\cite{Zhang_2021_synchro} while directionality may drive or hinder the synchronization~\cite{gallo2022synchro_hypergraphs}.
In consensus dynamics, \emph{effective graphs} exist only if the interaction between agents is a linear function~\cite{Neuhauser_2022_consensus_hypergraphs}.
The analysis of linear dynamics such as random walks on a hypergraph provides a framework to node ranking~\cite{carletti2020randomwalks, tran2019dir_hypergraph_pagerank, Bellaachia_2021} or node clustering~\cite{carletti2020stability, zhou2007learning_hypergraphs}.
Random walks on hypergraph can additionally be introduced artificially as an expedient to analyze the topology of the interactions~\cite{zhou2007learning_hypergraphs, Bellaachia_2021} or as a learning problem~\cite{Ducournau_2014, zhou2007learning_hypergraphs}.
Further, the necessity of analyzing such a system requires the development of new tools that extend and complete the measures developed for simple pairwise graphs and adapt them to the selected model~\cite{Estrada2006SubgraphCA, tudisco2021eigencentrality}.

In this work we leverage the results on directed hypergraphs~\cite{gallo1993directed,Ducournau_2014,Ausiello_2017,tran2019dir_hypergraph_pagerank} and recent findings characterizing dynamical systems on undirected hypergraphs~\cite{carletti2020stability,Eriksson2021flowbased}.
When the dynamics on the hypergraph are amenable to being described by a pairwise transition matrix, the latter defines an \emph{effective dynamics} on a pairwise topology that will microscopically reproduce the same (dynamical) behavior of the original higher-order graph.
We will show that, when the conditions are satisfied and the appropriate dynamics are defined on the hypergraph, a set of measures can be applied to them without the overhead of their eventual extension to the new framework.
Following this, one may expect that measures that are function of the dynamics, should return the same values on the hypergraph dynamics and its \emph{effective} pairwise counterpart, since the underlying transition matrix is identical.

In short, when analyzing a (possibly directed) hypergraph, we face the choice of extending the selected measure to the new framework, with the consequent overhead of increased complexity, or expand the hypergraph structure to a simple graph, with the consequent loss of information.
We propose a third route that involves measuring the dynamical process coupled to the hypergraph, in the form of its transition probabilities and steady state.

Finally, we characterize some real and synthetic examples to illustrate these findings.

\section{Directed hypergraphs and dynamical systems}%
\label{sec:Definitions}

While the choices of topological and dynamical models depend on each other, in the following, for simplicity, we describe them independently.

\subsection{Topology}%
\label{sub:Topology}

\begin{figure}[htb]
  \begin{center}
    \includegraphics[width=0.5\textwidth]{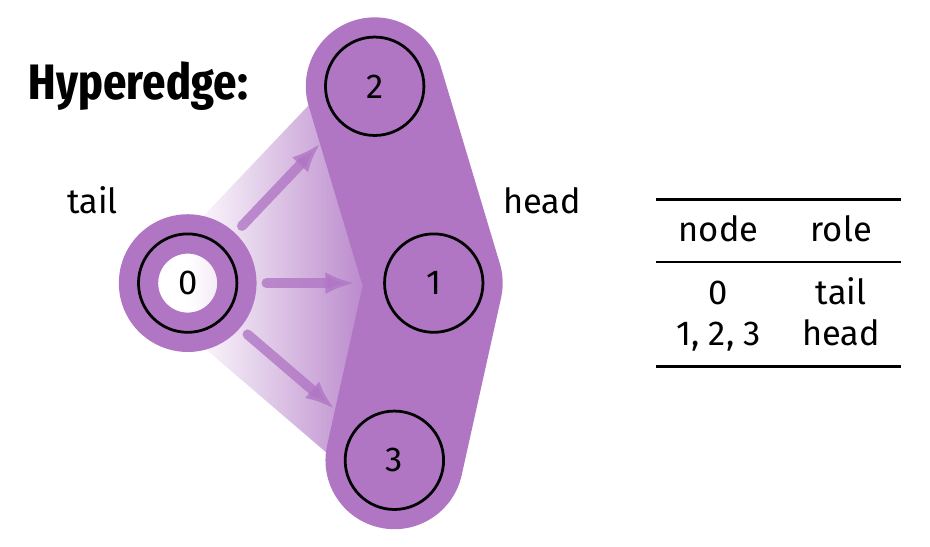}
  \end{center}
  \caption{%
    A hyperedge.
    Node $0$ is the \emph{tail} and represents the entry point of the hyperedge.
    Nodes $1, 2$ and $3$ are the exit points of the dynamics and are termed \emph{head}.
  }
  \label{fig:hyperedges}
\end{figure}

A hypergraph $G^h=\{N,E\}$ is defined by a set of nodes $N$ that interact through a set of hyperedges $E$.
A hyperedge $e \in E$ represents the interaction between a nonnegative number of nodes, although in the following we restrict it to the case where the interaction involves at least one node.
When the hyperedge is directed, the nodes involved in the interaction are divided into two non-empty (possibly overlapping) subsets: the source of the interaction is called the \emph{tail} of the hyperedge; and the target of the interaction is called the \emph{head}~\cite{Ausiello_2017} (see Fig.~\ref{fig:hyperedges}).
The nomenclature \emph{tail} and \emph{head} represent the directionality of the interaction through the figure of an arrow~\cite{Ausiello_2017}.

The boolean tail ($\mathsf{T}$) and head ($\mathsf{H}$) matrices are the extensions of \emph{incidence matrices}~\cite{zhou2007learning_hypergraphs} for the symmetric case, and are defined as follows:
\begin{align*}
  \mathsf{T}_{i\alpha} & =
  \begin{cases}
    1 & \text{if node $i$ is in the tail of }e_\alpha \\
    0 & \text{otherwise}
  \end{cases} \\
  \mathsf{H}_{i\alpha} & =
  \begin{cases}
    1 & \text{if node $i$ is in the head of }e_\alpha \\
    0 & \text{otherwise}.
  \end{cases}
\end{align*}
This construction is reminiscent of the chemical reaction hypergraphs~\cite{Mulas_2021} where incoming and outgoing interactions are defined by positive and negative values.

Hyperedge weights can be introduced as a weighting factor $w_\alpha$~\cite{carletti2020dynamical}, and define an interaction matrix $\mathcal{I}$:
\begin{equation}
  \mathcal{I} = \mathsf{T} W \mathsf{H}^\top,
  \label{eq:interaction_matrix}
\end{equation}
where $W = \diag(w)$ is the $|E|\times|E|$ weight matrix with hyperedge weights on the diagonal.
This matrix accounts for the weighted number of hyperedges that connect any ordered pair of nodes, in particular the weighted number of hyperedges where the tail encompasses the first node and the head encloses the second node.
This matrix recalls the adjacency matrix in graphs, and is often used as such in hypergraph expansion~\cite{Estrada2006SubgraphCA, Agarwal_2006, Li_1996_hypergraphs}.
While an adjacency matrix encodes all the information needed to reconstruct the topology, in the interaction matrix, part of that information is missing, in particular the higher order structure of the hyperedges.
We, therefore, strive to avoid this confusion and keep the two matrices conceptually separated.
Nevertheless, this partial information will be at the basis of the dynamical characterization in the next section.

\paragraph{Forward hyperedges}
In the following we suppose that all hyperedges have a tail of size 1, the so-called forward hyperedges~\cite{Ausiello_2017}.
If that was not the case, hyperedges with larger tails could decompose to a set of forward hyperedges each with the tail on a node of the former hyperedge and the same head.

In this framework, all columns of the tail matrix $\mathsf{T}$ have exactly one non-zero element ($\sum_i \mathsf{T}_{i\alpha} = 1\ \forall \alpha$).
We will refer to the size of the hyperedge ($|e|$) as the number of nodes included in its head: $|e_\alpha| = \sum_i \mathsf{H}_{i\alpha}$.
When all the hyperedges have size 1 ($|e_\alpha|=1\ \forall\alpha$), the hypergraph reduces to a directed graph with pairwise interactions.
A hypergraph is symmetric when, for any hyperedge $e_\alpha$, there exist other $|e_\alpha|$ hyperedges with a tail on each of its head nodes and a head on the other involved nodes (including the tail), with the same weight $w_\alpha$.

\paragraph{Hyperedge weights as function of the size.}

In some contexts, it may be useful to define the weight $w_\alpha$ of the hyperedge $e_\alpha$ as a function of the hyperedge size, in particular~\cite{carletti2020randomwalks}:
\begin{equation}
  w_\alpha = |e_\alpha|^\tau
  \label{eq:weight_tau}
\end{equation}
where $\tau$ is a bias parameter.
For positive values of the latter, hyperedges with large heads have larger weights, and vice versa for negative values.

\subsection{Dynamics on the hypergraph}%
\label{sec:dynamics}

Dynamical systems that leverage the hypergraph connectivity have been introduced several times in the literature~\cite{carletti2020randomwalks,Ducournau_2014,zhou2007learning_hypergraphs,Bellaachia_2021,Battiston_2020_hypergraphs_dynamics}.
A dynamical system linked to the hypergraph can be interpreted as a special case of the (hyper) graph signal processing~\cite{Schaub_Signal_proc_2021}, where the signal vector is constrained to be normalized and the shift operator is a right stochastic matrix.
In the following we describe a family of linear dynamics on a hypergraph, as defined in Ref.~\cite{carletti2020randomwalks} for the symmetric case.

Consider a dynamical system evolving on a directed hypergraph $G^h$ as a generalization of the classical random walk on a simple graph.
The transition probability that naturally follows from the interaction matrix of Eq.~\eqref{eq:interaction_matrix} can be written as:
\begin{equation}
  p_\tau(j|i)
  = \frac{\sum_{\alpha} \mathsf{T}_{i\alpha}\mathsf{H}_{j\alpha} w_\alpha^\tau}
  {\sum_{k\alpha} \mathsf{T}_{i\alpha}\mathsf{H}_{k\alpha} w_\alpha^\tau}
  = \frac{\mathcal{I}_{ij}(\tau)}{\sum_k \mathcal{I}_{ik}(\tau)}
  \label{eq:transition_prob}
\end{equation}
where $\tau$ is the bias parameter, and $\mathcal{I}$ the interaction matrix, with hyperedge weights as in Eq.~\eqref{eq:weight_tau}.
The above equation controls the walker dynamics which, sitting on node $i$, will perform the following steps:
\begin{enumerate}
  \item choose a neighboring hyperedge $e_\alpha$, whose tail includes node $i$, with a probability proportional to the hyperedge weight $p(\alpha|i) = |e_\alpha|^{\tau+1}$, and
  \item traverse the hyperedge to one of the nodes $j$ of its head, selected with even probability ($p(j|\alpha) = |e_\alpha|^{-1}$).
\end{enumerate}
This is a parametric family of biased random walks.
The bias parameter $\tau$ influences the dynamics warping the walker toward hyperedges with larger or smaller heads.
In particular, with $\tau=-1$ the dynamics will select each incident hyperedge with equal probability, independent of its size, similar to Ref.~\cite{zhou2007learning_hypergraphs}, while with $\tau=0$ it will select each incident hyperedge with a probability proportional to the respective hyperedge size.
The probabilities in Eq.~\eqref{eq:transition_prob} determine the entries of the right stochastic transition matrix $T_{ij}(\tau) = p_\tau(j|i)$.
In particular,
\begin{equation}
  T(\tau) = D^{-1}(\tau)\mathcal{I}(\tau)
  \label{eq:transition_interaction}
\end{equation}
where $D(\tau)=\diag(\mathcal{I}(\tau)\mathds{1})$ and $\mathds{1}$ is the vector of all ones.

In the case of ergodic dynamics, which implies the existence of a directed path between any two nodes, one can compute a steady state $\pi^\tau$ which encodes the probability of the walker to visit a given node of the hypergraph.
The steady state $\pi^\tau$ is the dominant eigenvector of the transition matrix $T(\tau)$ that corresponds to its unit eigenvalue.
In the following, we restrict the focus to ergodic dynamics.
Sometime some \emph{ad hoc} modifications are introduced to assure the ergodicity of an otherwise nonergodic system, e.g., the teleportation in the PageRank algorithm.

The probability of leaving node $i$ and reaching node $j$ at the following time step is $\Pi^\tau_{ij}=\pi^\tau_i p_\tau(j|i)$, and in matrix notation:
\begin{equation}
  \Pi^\tau = \diag(\pi^\tau) T(\tau) \,.
  \label{eq:probability}
\end{equation}
This matrix encodes all the information on the dynamical evolution of the system of interest, despite its higher order structure.
In the following we show how one can leverage the knowledge of the probability matrix (Eq.~\eqref{eq:probability}), the transition matrix (Eq.~\eqref{eq:transition_interaction}) and the steady state $\pi^\tau$ to characterize the hosting hypergraph.

\subsection{Hypergraph expansion}
\label{sec:expansion}

At this point, an interesting and probably na\"ive question is: is it possible to expand the hypergraph to a pairwise representation of it?
While an affirmative answer to this may question the usefulness of hypergraphs, several works have been done in this direction~\cite{Chaoqi2020hypergraphLineExpansion} in an effort to overcome the computation complexity introduced by the new framework.
A better question would be when and under which constraints we are allowed to expand the hypergraph to a pairwise representation.

Line or star expansion~\cite{Chaoqi2020hypergraphLineExpansion}, for instance, represents nodes and hyperedges as two classes of objects connected if incident in the original hypergraph (let us call them node-nodes and hyperedge-nodes respectively).
A random walk defined by such topology, when starting on a node-node, will jump to a neighboring hyperedge-node with even probability, and then move to another node-node chosen from the head of the hyperedge-node.
This dynamics reproduces the hypergraph dynamics defined by $T(\tau=-1)$ in Eq.~\eqref{eq:transition_prob} when we ignore all steps on hyperedge-nodes.

Another approach, introduced in graph learning~\cite{Chaoqi2020hypergraphLineExpansion, Ducournau_2014,  metz2021localization, Bellaachia_2021,Agarwal_2006}, is to define a random walk, derive from it the corresponding (pairwise) Laplacian and use it for, e.g., node labeling or transduction~\cite{Agarwal_2006}.
Although this approach may be effective and widely used in machine learning, the random walk is often introduced as an artifact to label nodes.

Can we find an \emph{effective graph} or adjacency matrix for which a random walk reproduces the expected transition probabilities and steady state distribution?

\paragraph{Symmetric hypergraphs.}
When considering a symmetric hypergraph, the interaction matrix $\mathcal I(\tau)$ is also symmetric, as it represents the (possibly weighted) count of hyperedges incident to any two nodes $i$ and $j$.
As on a symmetric graph, the steady state distribution will be:
\begin{equation*}
  \pi^\tau =
  % \frac{\mathds{1}^\top \mathcal{I}(\tau)}{\|\mathcal{I}(\tau)\|} =
  \frac{\mathcal{I}(\tau)\mathds{1}}{\|\mathcal{I}(\tau)\|}
\end{equation*}
where $\|\cdot\|$ is the $L_{1,1}$ norm.
In fact, one can see that the above is a stationary state:
\begin{equation}
  \pi_j = \sum_i \underbrace{\frac{\mathcal{I}_{ij}}{\sum_k \mathcal{I}_{ik}}}_{p(j|i)} \cdot \underbrace{\frac{\sum_k \mathcal{I}_{ki}}{\|\mathcal{I}\|}}_{\pi_i}
  =
  \sum_i \frac{\mathcal{I}_{ij}}{\|\mathcal{I}\|}\,.
\end{equation}
where we omitted the dependence on $\tau$.
Finally,
\begin{equation*}
  \Pi^\tau =
  \diag(\pi^\tau) T(\tau) =
  \frac{D(\tau)D^{-1}(\tau)\mathcal{I}(\tau)}{\|\mathcal{I}(\tau)\|} =
  \frac{\mathcal{I}(\tau)}{\|\mathcal{I}(\tau)\|}\,.
\end{equation*}
The above entails that there exists a family of \emph{effective adjacency matrices} that are all proportional to the interaction matrix and reproduce the same random walk dynamics.

Unfortunately, this is not the case for a directed hypergraph.
There exists, in fact, a (infinite) family of graphs that correspond to a given transition matrix.
Any matrix of the form $A = \diag(\vec v) T(\tau)$, for any non-negative vector $v$, is a suitable adjacency matrix, and a natural choice does not exist.
In particular, the interaction matrix $\mathcal{I}(\tau)$ (with $\vec v = \mathcal I(\tau)\mathds 1$), the probability matrix $\Pi^\tau$ ($\vec v = \pi^\tau$), and the transition matrix itself ($\vec v = \mathds 1$) are three representatives of that family.
These adjacency matrix candidates display different linking patterns and a choice between them cannot be done \emph{a priori}.

\paragraph{Clique expansion.}
Often, in the literature, one can find approaches involving the \emph{clique expansion} of symmetric hypergraphs~\cite{Agarwal_2006, Zien96multi-levelspectral, De_Castro_1999_erdos_number, Newman_2001}.
Here, a hyperedge is replaced by a clique of (pairwise) links between all the nodes involved in the original higher-order interaction.
This approach assumes that the dynamics can freely flow between any pair of nodes involved in the interaction.

Similarly to the case of symmetric hypergraphs, \emph{clique expansion} uses the interaction matrix of Eq.~\eqref{eq:interaction_matrix} as adjacency matrix with $\tau=0$.
This choice imposes a precise linear dynamical model where a random walker will choose neighboring hyperedges according (proportionally) to their head size.

\section{Measuring dynamics}
\label{sec:measures}

In the previous section, we characterized the dynamical process evolving on the hypergraph and analyzed when an effective adjacency matrix that reproduces the same dynamics can be written.
In the following we abandon the quest for an effective adjacency matrix and, instead, we propose to measure the dynamical system itself.

One may expect that measures that are mere functions of the dynamical evolution should return the same result when applied to a dynamical system on a hypergraphs or to a pairwise graph, when those display the same dynamical behavior.
Measures of the transition matrix $T(\tau)$ (Eq.~\eqref{eq:transition_prob}), the probability matrix $\Pi^\tau$ (Eq.~\eqref{eq:probability}) or the steady-state probability distribution $\pi^\tau$, that characterize the dynamical system, will depend on the value of the biasing parameter $\tau$.
However, those measures involve pairwise matrices and provide a basis for a straightforward extension of similar measures of simple graphs.

\subsection{Node ranking}
\label{sec:ranking}

For a simple graph, the PageRank~\cite{brin1998pagerank} $x_i$ of node $i$ represents its level of connection to other highly ranked nodes.
This algorithm was introduced to rank web-pages according to their linking pattern and can be mapped to the problem of finding the steady state of a random walker with teleportation.
\begin{equation}
  x \left( (1-\alpha) T(\tau) + \frac{\alpha}N \mathds 1_{NN}\right) = x
  \label{eq:pagerank}
\end{equation}
where $N$ is the number of nodes, and $\mathds 1_{NN}$ is the $N\times N$ matrix of all ones.
On the left side of Eq.~\eqref{eq:pagerank}, the first term represents the dynamical evolution under the transition matrix $T(\tau)$ and the second term introduces a teleportation to any other node of the network which happens with probability $\alpha$.
This term assures the ergodicity of the dynamics.

\begin{figure}[htpb]
  \centering
  \includegraphics[width=\linewidth]{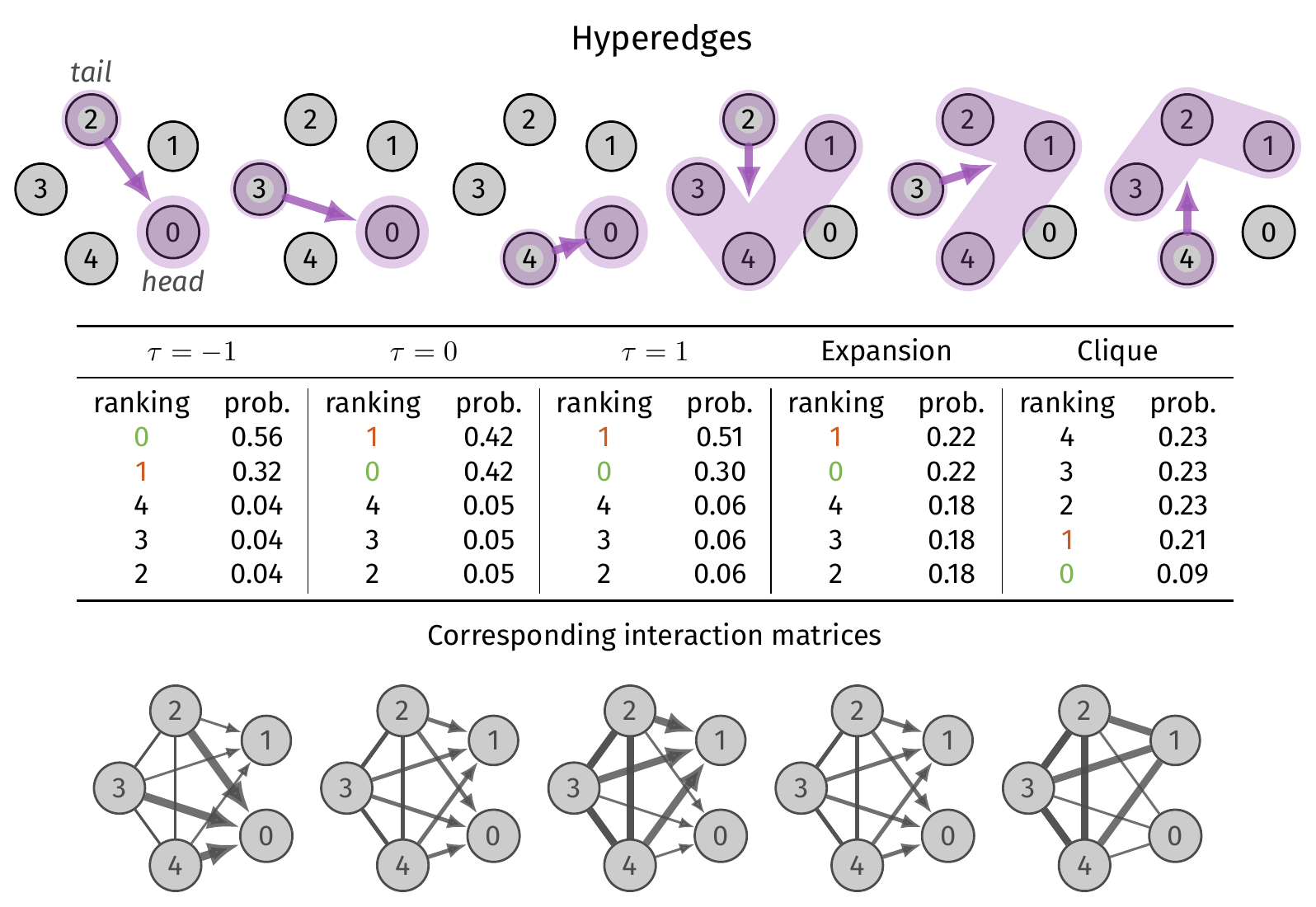}
  \caption{%
    Example of a hypergraph of 5 nodes with 6 directed hyperedges (listed on top).
    The PageRank node ranking changes depending on the value of the bias parameter $\tau$ (See table).
    The interaction matrix corresponding to the three values of $\tau=\{-1, 0, 1\}$ are represented on the lower part of the plot.
    One can notice, in particular, that node 0 and node 1 change their relative order within the ranking.
    For comparison, ranking and adjacency matrix of the hypergraph expansion (where the adjacency matrix corresponds to the interaction matrix $\mathcal I(\tau=0)$) and of the clique expansion are reported.
  }%
  \label{fig:eigencentrality}
\end{figure}

Solving Eq.~\eqref{eq:pagerank} only involves the knowledge of the transition matrix which enable us to compute the PageRank for the hypergraph.

In Fig.~\ref{fig:eigencentrality} we display a didactic example of a hypergraph of 5 nodes and 6 hyperedges.
In this graph, hyperedges of size 1 point to node 0, while hyperedges of size 3 point to node 1 but never to node 0.
Note how the ranking of nodes, according to the PageRank defined in Eq.~\eqref{eq:pagerank}, depends on the value of the bias parameter $\tau$.
For negative values of $\tau$ the dynamics follow preferentially small edges, hence node 0 has the highest rank.
For positive values of $\tau$ the highest ranked node is node 1.

A hypergraph expansion using the interaction matrix $\mathcal I(\tau=0)$ as the adjacency matrix will expand each hyperedge to a set of edges from the tail to each node of the head, with weight 1.
Fig.~\ref{fig:eigencentrality} reports the ranking according to this expansion and highlights that, while the node ordering is the same as for $\tau=0$, the difference between high and low probabilities is much lower.
A clique expansion, ignoring the directionality of hyperedges, leads to the most central nodes of Fig.~\ref{fig:eigencentrality} being $\{2, 3, 4\}$, followed by node $1$ and a way less central node $0$.
This expansion completely inverts the ordering of the nodes dictated by the dynamics.

Tran \emph{et al.}, in Ref.~\cite{tran2019dir_hypergraph_pagerank}, use a similar approach fixing the bias parameter $\tau=-1$.
However, they use a symmetrized version of the Laplacian instead of computing the dominant eigenvector from the transition matrix.

Tudisco \emph{et al.}~\cite{tudisco2021eigencentrality} introduce a different approach to calculate the eigencentrality of a hypergraph inspired by the Hyperlink-Induced Topic Search HITS algorithm.
Here, the importance of nodes and edges depends on the importance of the neighboring edges and nodes respectively (as opposed to using hub and authority centralities for nodes only).

\subsection{Community detection}
\label{sub:community}

Extending community detection algorithms that involve a measure of the dynamics is also within reach.

\begin{figure*}[!t]
  \begin{center}
    \includegraphics[width=\textwidth]{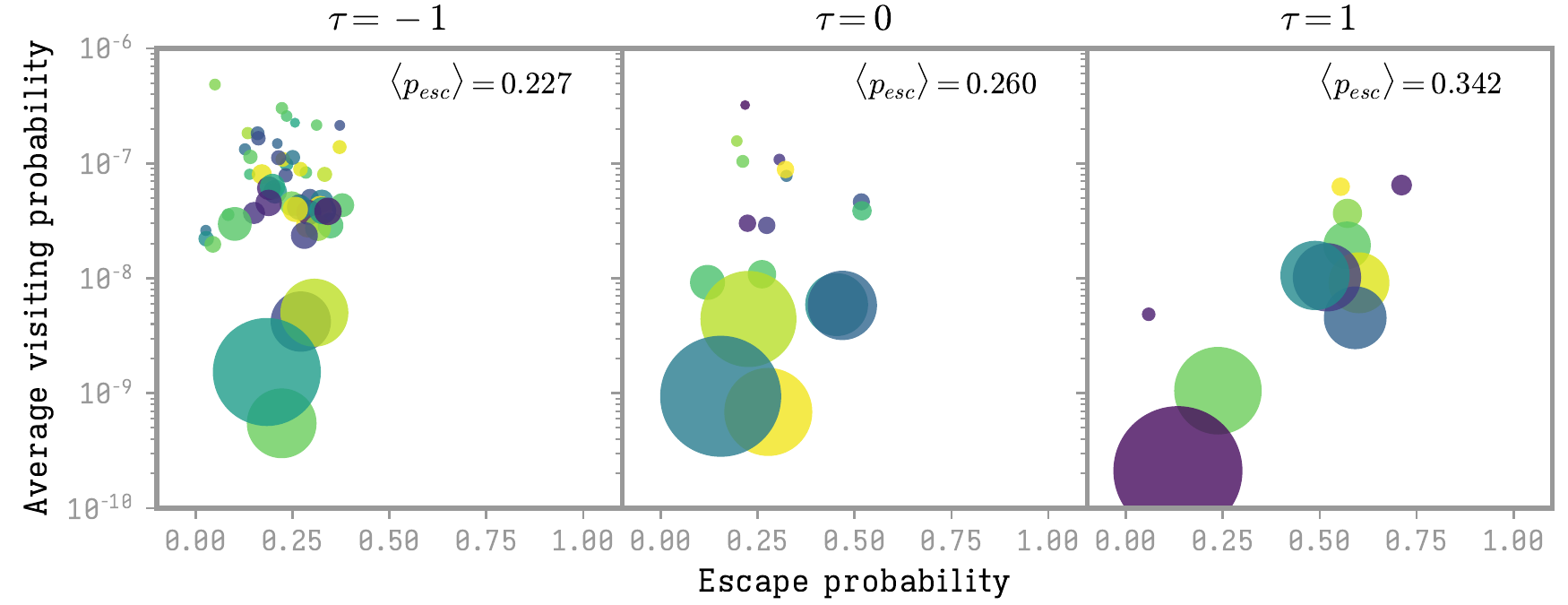}
  \end{center}
  \caption{%
    Communities of the French-speaking Twitter, discussing vaccines during the COVID-19 pandemic.
    Probability that a random walker on a community exits its boundaries (escape probability) and the average probability that the walker visit a node of the community (average visiting probability) for three values of $\tau\in\{-1, 0, 1\}$.
    The shift of prolific communities (high average visiting probability) toward the right, when $\tau$ increases, shows that the information that reaches different communities comes mostly from highly retweeted tweets.
    Communities are computed on $\Pi^\tau$ and their size is proportional to the number of users.
  } \label{fig:tweets}
\end{figure*}

Modularity~\cite{newman2004finding} is a network measure that, for a proposed community structure, accounts for the amount of connectivity within communities, compared to a random null model.
Its maximization is the basis of several widely used algorithms~\cite{blondel2008modularity,traag_leiden_2019,Delvenne2010stability}.
In a symmetric graph, the original modularity form is
\begin{equation}
  Q = \frac{1}{2m} \sum_{ij} \left(A_{ij} - \frac{k_ik_j}{2m}\right) \delta_{c_ic_j},
  \label{eq:modularity}
\end{equation}
where $A$ is the adjacency matrix, $k_i=\sum_j A_{ij}$ is the node degree, $m=\frac 12 \sum_{ij}A_{ij}$ is the number of edges and $c_i$ is the community assigned to node $i$.

While Eq.~\eqref{eq:modularity} is a function of the linking pattern, it has been shown that it can be rewritten as a sum of autocovariances of a random walk~\cite{Delvenne2010stability, Schaub2019multiscale}:
\begin{equation}
  Q = \sum_c \text{cov}\left(\chi_c(t), \chi_c(t+1)\right)
  \label{eq:covariance}
\end{equation}
where $\chi_c$ is the characteristic function of partition $c$.
This provides a dynamical interpretation of modularity maximization, which accounts for the trapping effect of a community on the random walk dynamics.
Leicht and Newman extend the modularity function to directed networks~\cite{Leicht2008CommunitySI} disentangling the modularity definition from its dynamical interpretation.
In fact, the above relation between modularity and auto-covariance holds for the symmetric case.

Inspired by the relation between covariance and modularity, Delvenne \emph{et al.}~\cite{Delvenne2010stability} provide an extension of Eq.~\eqref{eq:covariance} to directed graphs and to different time scales.
Here, the covariance form of modularity can be rewritten as:
\begin{equation}
  Q(\tau) = \sum_c \sum_{ij\in c} p^\tau(i,j) - \pi^\tau_i\pi^\tau_j
  \label{eq:hyper_cov}
\end{equation}
that only depends on $\Pi^\tau$, which assumes the role of adjacency matrix.

This formulation, which utilizes the dynamical system to cluster the graph nodes, let us extend the community detection algorithm to hypergraphs~\cite{carletti2020stability}.

\begin{figure}[!t]
  \begin{center}
    \includegraphics[width=\linewidth]{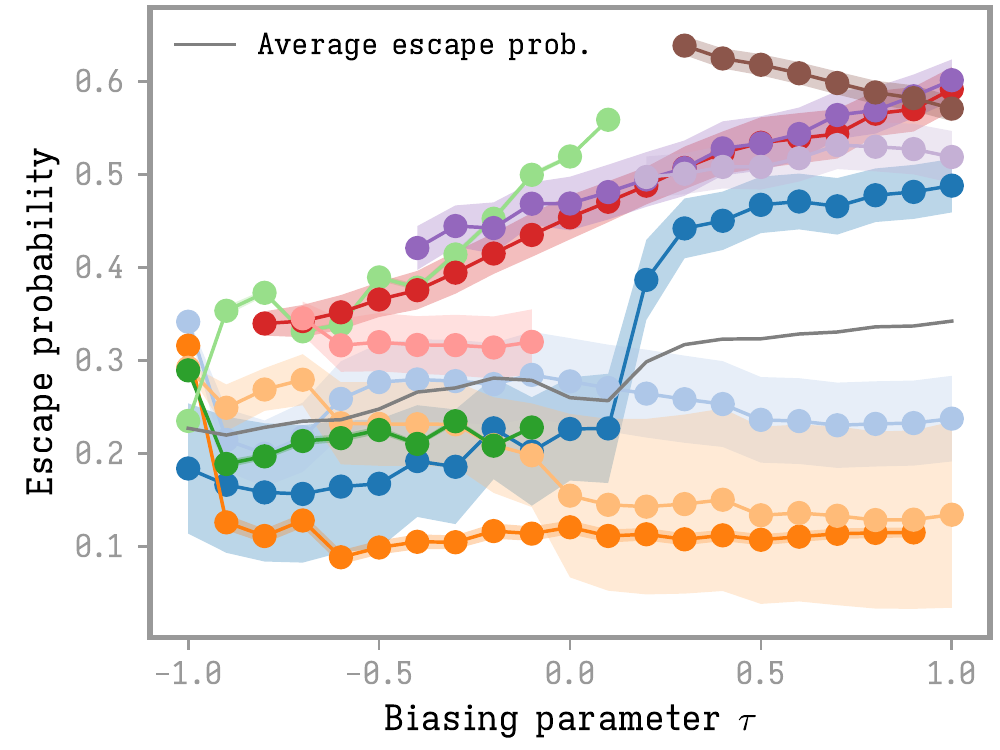}
  \end{center}
  \caption{%
    Escape probability per community as a function of $\tau$ (the gray line represents the global average), with the same data as in Fig.~\ref{fig:tweets}.
    Note that most of the communities increase their capacity to reach other communities as the value of $\tau$ increases.
    Fluctuations are often due to the splitting or merging of communities as $\tau$ changes.
    The filled area around the lines is proportional to the number of nodes in each community.
    For readability only communities with more than 1000 users are reported.
  } \label{fig:community_escape}
\end{figure}

An online social network like Twitter, where the discussion follows a one-to-many pattern, represents a system where directed hypergraphs are best suited.
Fig.~\ref{fig:tweets} shows the community structure of users on the French-speaking part of Twitter, discussing vaccine related topics in the context of the COVID-19 pandemic (dataset from Ref.~\cite{faccin2022covid}) between 2020-10-01 and 2020-12-31 (the period in which the first COVID-19 vaccines became available).
Each tweeting user represents the tail of a hyperedge whose head nodes are the retweeting users.
Communities are detected through the maximization of the auto-covariance using the \texttt{python} implementation of the stability algorithm: \texttt{PyGenStability}~\cite{PyGenStability}.
To assure ergodic dynamics, for each forward interaction we add a weaker backward interaction:
\begin{equation*}
  \mathcal{\tilde I}(\tau) = \mathcal I(\tau) + \delta\mathcal I^\top(\tau)
\end{equation*}
with $\delta=0.1$, and consider the largest strongly connected component.

In Fig.~\ref{fig:tweets} we plot, for different values of $\tau$, the average visiting probability (the probability of visiting a node of a given partition, on average):
\begin{equation*}
  p_\text{visiting}(c) = \sum_{i\in c} \frac{\pi^\tau_i}{|c|}\,,
\end{equation*}
and the escape probability (the probability of reaching a node on a different community) of each community:
\begin{equation*}
  p_\text{escape}(c) = \frac{\sum_{i\in c, j\notin c} \Pi_{ij}^\tau}{\sum_{i\in c} \pi^\tau_i}\,.
\end{equation*}
When comparing the communities computed with $\tau=-1$ (each tweet has the same weight) or $\tau=0$ (each tweet weight is proportional to the number of its retweets) one can notice that the in the upper part of the plots (the most prolific communities), there is a shift toward higher escape probabilities.
The dynamics corresponding to higher values of $\tau$ favors hyperedges with larger heads (larger number of retweets), hence we may conclude that tweets with larger retweet cascades have a higher chance to link to users of different communities.

To show this, Fig.~\ref{fig:community_escape} displays the escape probability of the larger communities at varying values of $\tau$.
Note that the probability of reaching nodes on different communities is an increasing function of $\tau$ in the range of values analyzed.
The communities are recomputed independently for each value of $\tau$ and communities of successive $\tau$ steps, that overlap for more than half of the users, are merged to form a time evolving community.

\begin{figure}[htb]
  \begin{center}
    \includegraphics[width=\linewidth]{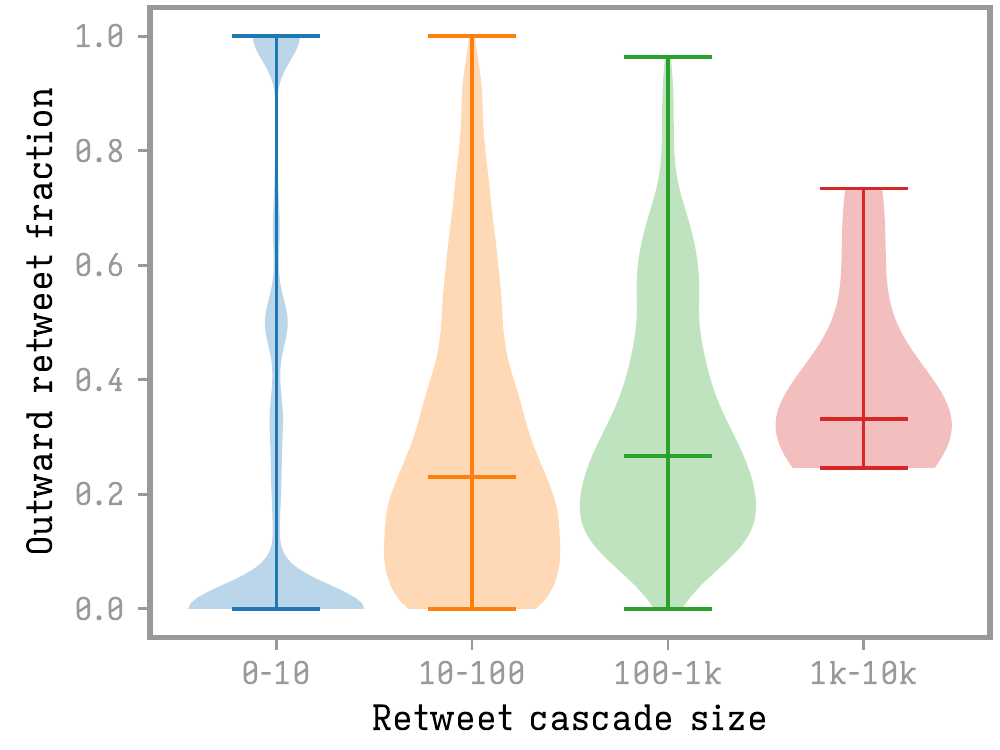}
  \end{center}
  \caption{%
    Tweet capacity to reach outer communities.
    Proportion of retweets from other communities as a function of the retweet-cascade size for the community structure detected at $\tau=-1$.
    Results are robust upon changes to the value of $\tau$.
  }
  \label{fig:out_retweet_cascade}
\end{figure}

To further support this observation, we analyze the linking pattern of the hypergraph.
In particular, we study the capacity of tweets of different cascade size to reach a community other than that of the original poster.
Figure~\ref{fig:out_retweet_cascade} supports this hypothesis showing the ratio of retweets reaching other communities per tweet class.
Here one can notice the increasing probability of being retweeted outside of one's community when the tweet become more and more popular.

Modularity of the detected community structure (Eq.~\eqref{eq:hyper_cov}) monotonically decreases from 0.70 ($\tau=-1$) to 0.36 ($\tau=1$).
This modularity behavior may be influenced by the linking pattern: negative values of $\tau$ bias the walker toward small hyperedges, which have a higher chance of connecting nodes of the same community.
Applying the same community detection algorithm to the retweet network which corresponds to the interaction matrix $\mathcal I(\tau=0)$, one finds a partition with a higher number of small communities that fall closer to that corresponding to $\tau=0$ (with normalized mutual information 0.49).
Further, the community structure detected in the latter case displays a lower modularity $Q=0.40$ than the corresponding modularity detected through Eq.~\eqref{eq:hyper_cov} for most of the considered values of $\tau$ (in particular for $\tau\le 0.5$).

\paragraph{Other algorithms.}
Within the vast ecosystems of community (or other mesoscopic structures) detection algorithms, those for which the objective function only depends on the system dynamics, can be extended as in this work.
Some notable examples are the map-equation~\cite{Rosvall_2009} which was extended to symmetric hypergraphs in Ref.~\cite{carletti2020stability}.
Another algorithm based on information theoretical arguments amenable to extension is the autoinformation state aggregation~\cite{faccin2021aisa}.

\section{Discussion}%
\label{sec:Discussion}

We discussed how the analysis of higher order graphs can be performed through analysis of the dynamical system coupled to the hyperedge topology.
When a transition matrix can be computed for the dynamics, all measures that are function of the latter can be applied directly.
One can assume that a function of the dynamics has the same results if dynamics on a hypergraph and on the corresponding graph have identical transition matrices.

Prudence is necessary.
Although an \emph{effective} transition matrix may exist, to the latter corresponds a whole family of possible adjacency matrices which produce the same transition probabilities.
While in the symmetric case one can na\"ively analyze the interaction matrix of Eq.~\eqref{eq:interaction_matrix} as the adjacency matrix of an effective graph, this can lead to misconceptions.
In particular, one may be tempted to analyze the linking pattern of such an effective graph incurring an improper extension of the measure of interest.
In the directed case, the analysis of the interaction matrix does not directly relate to the dynamical behavior of the original system.

Other popular approaches involve an expansion of the hyperedge pattern, e.g., clique expansion, which assumes a specific dynamical model that may not match the original system.
If the interest of the researcher is instead in the linking pattern of the hypergraph, one must redefine the measure of interest, as in Ref.~\cite{tudisco2021eigencentrality, Estrada2006SubgraphCA} where the authors define an eigencentrality or clustering coefficient adapted to this framework.

Rather than expanding the hypergraph to an effective adjacency matrix that may involve the mentioned drawbacks, or the extension of the measure to the hypergraph framework, we propose to measure the dynamical process itself, when possible.

In this work we focused on dynamical systems inspired by random walks, but other kinds of dynamics can also be investigated~\cite{carletti2020dynamical}.

A topic of interest for further studies would be how to determine the dynamical model to use on the hypergraph structure, in particular, in the linear case, the best value of $\tau$.
In many cases, as in the present work, the choice falls on the researcher, and they need to determine the value of $\tau$ or to analyze a range of values.
Another way would be to use one of the many model selection approaches applied to the measure of interest, such as the elbow or the plateau criterion.

Another topic of interest for further studies is determining when this framework is applicable to nonlinear dynamics such as non-Markovian dynamics~\cite{salnikov2016} or quantum mechanical systems~\cite{biamonte2019classicalquantum, Rossi_2013}.
In the former the dynamics depends on the symbols visited in the past (the dynamics is linear on the set of present and past symbols).
In the latter the quantum system evolves on the nodes, and, while there is no steady state, one may consider temporal averages or open quantum systems (coupling the system to a bath).

\acknowledgments

MF thanks Michael T. Schaub for the insightful discussions on the topic.
Part of this work has benefited from the financial support of the National French Agencies ANR (project TRACTRUST - ANR-20-COVI-0102) and ANRS (project MEDIACAM - ANRS COV24).

\bibliography{biblio}
\bibliographystyle{apsrev4-2}
\end{document}